\documentclass[11pt]{JHEP3}

\usepackage{amssymb,amsmath}
\usepackage{epsfig}
\usepackage{amsfonts}
\usepackage{graphicx}
\usepackage{amscd}
\usepackage{dsfont}

\newtheorem{lemma}{Lemma}[section]

\newtheorem{theorem}{Theorem}[section]
\newtheorem{proof}{Proof}[section]
\newtheorem{definition}{Definition}[section]

\title{Open and Closed String Worldsheets from Free Large $N$ Gauge Theories with Adjoint and Fundamental Matter}
\author{Itamar Yaakov\\Department of Particle Physics,\\
Weizmann Institute of Science,\\Rehovot 76100,
Israel.\\\email{itamar.yaakov@weizmann.ac.il}} \abstract{ We extend
Gopakumar's prescription for constructing closed string worldsheets
from free field theory diagrams with adjoint matter to open and
closed string worldsheets arising from free field theories with
fundamental matter. We describe the extension of the gluing
mechanism and the electrical circuit analogy to fundamental matter.
We discuss the generalization of the existence and uniqueness
theorem of Strebel differentials to open Riemann surfaces. Two
examples are computed of correlators containing fundamental matter,
and the resulting worldsheet OPE's are computed.
Generic properties of Gopakumar's construction are discussed.}
\preprint{WIS/08/06-JULY-DPP}

\begin{document}

\section{Introduction and review}

In this work we extend a prescription proposed by R. Gopakumar
\cite{Gopakumar:2003ns,Gopakumar:2004qb,Gopakumar:2005fx,Gopakumar:2004ys},
for connecting free large $N$ $U(N)$ quantum gauge theories with
adjoint matter fields and closed string theories, to field theories
containing also fundamental matter fields and string theories
containing also open strings. The construction gives a suggestion
for the correlation functions on the string worldsheet of the string
theories dual to free large $N$ gauge theories, which may be
relevant in the context of the AdS/CFT correspondence
\cite{Maldacena:1997re}. The prescription involves matching the
moduli space of Schwinger parameters of Feynman diagrams on the
field theory side with that of Riemann surfaces in the string
theory. In the rest of this section we briefly review Gopakumar's
construction.

In section \ref{prelimenaries} we extend the prescription to
worldsheets with boundary. The gluing construction for correlators
with fundamental matter is discussed. We prove a simple extension
of Strebel's theorem for quadratic differentials to Riemann
surfaces with boundary. We outline our method for constructing
differentials on such surfaces.

In sections \ref{three} and \ref{four} we compute two explicit
examples using the generalized prescription, and the results are
discussed. The worldsheet OPE is examined. We elaborate on some
general aspects of the construction. The appendix includes an
analysis of the worldsheet OPE for a more general class of diagrams.

\subsection{Gopakumar's prescription}

We summarize a prescription due to Gopakumar \cite{Gopakumar:2005fx}
for implementing a duality between a free large $N$ field theory
with matter in the adjoint representation of the gauge group and an
unknown closed string theory. Each Feynman diagram contributing to a
specific correlation function is mapped to an integral over the
moduli space of marked Riemann surfaces, and the integrand is
interpreted as a correlation function in the dual worldsheet CFT. We
will use the coordinate space version of the Schwinger
parametrization for field theory correlators, and discuss only the
simplest case of diagrams involving massless scalar fields in four
dimensions (the generalization to other cases is straightforward).

Each free field theory diagram is a product of propagators of the
form $1/(x_i-x_j)^2$, and in the Schwinger parametrization we
rewrite this as
$\frac{1}{\left(x_{2}-x_{1}\right)^2}=\int_{0}^{\infty}d\sigma
e^{-\sigma\left(x_{2}-x_{1}\right)^2}$.

When the diagram has several propagators which are homotopic (as
lines on a Riemann surface, when we draw the Feynman diagram in 't
Hooft's double line notation), we can glue them together. The
diagram then depends only on the sum of the Schwinger parameters,
and we can write a ``glued" propagator in the form
$\frac{1}{\left(x_{2}-x_{1}\right)^{2n}} \propto \int
d\tilde{\sigma}
 \tilde{\sigma}^{n-1}
 e^{-\tilde{\sigma}\left(x_{2}-x_{1}\right)^2}$.
A diagram in which all homotopic edges have been glued is known as a
skeleton graph. An $n$-point function is generated by a sum over the
relevant skeleton graphs and over the multiplicities of their edges.

In order to translate to Riemann surfaces we use quadratic Strebel
differentials. These are tensors $q=\phi(z)(dz)^2$ defining an
invariant line element $ds=\sqrt{\phi(z)}dz$ on a Riemann surface. A
straight arc, $\gamma(t)$, of a differential is one that satisfies
$\arg(\phi(\gamma(t))(d\gamma/dt)^2)=\theta$, with the values
$\theta=0$ and $\theta=\pi$ defining horizontal and vertical curves,
respectively. Strebel differentials are a particular class of
quadratic differentials which solve a minimal area problem. We
define a Strebel differential by the following theorem due to K.
Strebel \cite{Strebel:1980} :

\begin{theorem}
\label{type1} Given an n-punctured genus $g$ Riemann surface $R$
($g\geq 0$, $n\geq 1$ and $2-2g-n<0$), with prescribed locations
$z_{i}$ for the punctures, and a set of positive numbers
$\{p_{i}\}_{i=1}^{n}$, there exists a unique quadratic differential
$q$ on $R$ satisfying the following conditions:
\begin{itemize}
\item $\phi(z)$ is meromorphic on $R$, and its only poles are
double poles at the locations of the punctures. The residue at the
$i$'th pole is $p_{i}$, in the sense that $\frac{1}{2\pi
i}\oint_{\gamma_{i}^{\alpha}}\sqrt{\phi(z)}dz=p_{i}$ for all
appropriate curves; integrations are performed in the direction
that makes $p_{i}>0$. \item The non-closed horizontal trajectories
are of measure zero on the Riemann surface.
\end{itemize}
\end{theorem}

The closed horizontal trajectories foliate ring domains centered
at the locations of punctures (a numerical demonstration is
available in \cite{Moeller:2004yy}). The non-closed horizontal
trajectories, also known as critical curves, connect the various
zeros of the function $\phi(z)$. These curves describe a graph
which is embedded into the Riemann surface. We can associate a
length $l_{i}$ with each edge of this graph, the length of the
corresponding curve as measured by the line element $ds$, thus
constructing a metric graph. The space of all ribbon graphs
(another name for the double line type graphs of the large $N$
expansion) with length assignment to each edge and all $n$
vertices of order three or more is known as ${\cal
M}^{comb}_{g,n}$ \cite{Kontsevich:1992ti}.

The space of genus $g$ Riemann surfaces with $n$ marked points and
a positive number $p_{i}$ assigned to each point is known as the
decorated moduli space ${\cal M}_{g,n}\times \mathbb{R}_{+}^{n}$.
It turns out that ${\cal M}^{comb}_{g,n}$ gives a cell
decomposition of ${\cal M}_{g,n}\times \mathbb{R}_{+}^{n}$
\cite{Gopakumar:2005fx,Kontsevich:1992ti}. We can then show a one
to one correspondence between the space of Strebel differentials
and this space. The ribbon graph of the field theory correlator
gives a triangulation of the surface.

Gopakumar's prescription consists in identifying the conductances
$\sigma_{i}$ of the field theory skeleton graph with the lengths of
the corresponding Strebel differential: $\sigma_{i}\equiv l_{i}$.
One must then integrate over the $\mathbb{R}_{+}^{n}$ factor of the
decorated space to obtain an expression which depends only on the
moduli of the Riemann surface. Thus, every Feynman diagram is
rewritten as an integral over the moduli space of a Riemann surface,
which is interpreted as a correlation function in the dual string
theory. Note that the moduli count on the two sides of the
correspondence is the same. On the field theory side this is the
number of edges, while on the worldsheet it is the number of real
moduli together with a positive number associated with each vertex
operator.

\section{Generalization to fundamental representation fields/open strings}
\label{prelimenaries}
 In this section we describe the
generalization of Gopakumar's prescription to field theories
containing matter in the fundamental representation of the gauge
group. The Feynman diagrams in such theories will correspond to open
string worldsheets, with boundaries on the fundamental
representation propagators. Such worldsheets are Riemann surfaces
with boundaries and punctures. Punctures may occur both in the
interior of the surface (for operators involving only adjoint
fields) and on the boundaries (for operators bilinear in fundamental
representation fields). We describe how the gluing mechanism works
for graphs generated by these Feynman diagrams. The theorem
regarding Strebel differentials is extended to include the case of
Riemann surfaces with boundaries. Finally we describe our method of
constructing the differentials for such surfaces using image
charges.

\subsection{The gluing mechanism and fundamental propagators}

The gluing construction described in
\cite{Gopakumar:2004qb,Gopakumar:2005fx,Bjorken:1965} can be
carried over to the case of correlators with fields in the
fundamental representation. The point to keep in mind is that the
gluing of lines must respect the color flow prescribed by the
contractions of the correlator. Specifically it cannot change the
nature of the two dimensional surface that the double-line graph
describes which, in the large $N$ expansion, is related to the
order of the correlator in the string theory perturbative
expansion. Geometrically the lines in the graph corresponding to
propagation of particles in the fundamental representation
describe a fixed boundary. Adjoint matter lines in the interior
may be homotopic to a boundary line or to each other, in which
case they may be glued.

\subsection{Strebel differentials on Riemann surfaces with boundary}

The theory of Strebel differentials described in the introduction
deals with compact Riemann surfaces. This is appropriate for
closed string worldsheets. We would like to extend Gopakumar's
prescription to open + closed string worldsheets. To this end we
need a generalization of the cell decomposition provided by
Strebel differentials and ribbon graphs to Riemann surfaces with
boundary. Such a generalization already exists
\cite{Zwiebach:1990ba} for specific differentials. We summarize
briefly the properties of this construction and prove the
necessary extension. We begin with some definitions
\cite{Strebel:1980}:

\begin{definition}
\label{Riemann} A Riemann surface $R$ is a connected Hausdorff
space $M$ together with an open covering $\{U_{\nu}\}$ and a
system of homeomorphisms $h_{\nu}$ of the sets $\{U_{\nu}\}$ onto
open sets $V_{\nu}=h_{\nu}(U_{\nu})$ in the complex plane
$\mathbb{C}$ with conformal neighbor relations. A bordered Riemann
surface is defined by homeomorphisms to the closed half plane
(which we choose to be the upper half-plane). The set of points
mapping to the real line, denoted $\Gamma$, is the border of the
Riemann surface. Note that $\Gamma$ is a one dimensional manifold
that is not necessarily connected. Every connected component of
$\Gamma$ is a border of $R$.
\end{definition}

\begin{definition}
Let $R=(M,\{(U_{\nu},h_{\nu})\})$. The mirror image of $R$ is
defined to be the surface $R^{*}=(M,\{(U_{\nu},\bar{h_{\nu}})\})$.
Every Riemann surface, bordered or not, has a mirror. Note that,
with our conventions, coordinates for the mirror live in the lower
half plane.
\end{definition}

\begin{definition}
The double $\hat{R}$ of a bordered Riemann surface $R$ is the
union of $R$ and $R^{*}$ with the points on $\Gamma$ identified.
This turns out to be a legitimate Riemann surface with the mapping
to the complex plane defined either by $h_{\nu}$ or
$\bar{h}_{\nu}$ depending on whether the point in question was in
$R$ or $R^{*}$. The two mappings naturally agree on the set of
identified points $\Gamma$. Note that the base manifolds for $R$
and $R^{*}$ are the same; thus every connected component of
$\Gamma$ has a unique mirror image. The doubling identifies the
original connected component and its mirror and is therefore
unambiguous.
\end{definition}

\begin{theorem}
\label{type2}  For every punctured Riemann surface $R$ with boundary
$\partial R$, with prescribed positions for the $n$ punctures, some
of which may be on the boundary, and a set of positive numbers
$\{p_{i}\}_{i=1}^{n}$ there exists a unique quadratic differential
$q=\phi(z)d^2z$ possessing all the properties listed for Theorem
\ref{type1} and additionally:
\begin{itemize}
\item $\phi(z)$ is real on the boundary. \item The residues of
poles on the boundary of $R$ are $\frac{1}{\pi
i}\oint_{\gamma_{i}^{\alpha}}\sqrt{\phi(z)}dz=p_{i}$, where
$\gamma_{i}^{\alpha}$ are again curves homotopic to the puncture
on the boundary. These begin and end on the two boundary segments
separated by the puncture (these lie in the same connected
component of $\Gamma$).
\end{itemize}

\end{theorem}

\begin{proof}
Let $\hat{R}$ be the double of the surface $R$. Let $q$ be the
Strebel differential of Theorem \ref{type1} on $\hat{R}$ with the
residues $\{p_{i}\}_{i=1}^{n}$ specifying the residue of both a
puncture and its mirror image.
\begin{lemma}
$q$ is invariant under the anti-holomorphic automorphism on
$\hat{R}$ that exchanges $R$ and $R^{*}$.
\end{lemma}
\begin{proof}
The image of $q$ under this automorphism is also a quadratic
differential $\tilde{q}$. It is easy to see that $\tilde{q}$
satisfies all the demands of Theorem \ref{type1}, therefore
$q\equiv\tilde{q}$. In particular, $q$ is real on $\Gamma$. Note
that the reality condition holds for every connected component of
$\Gamma$. The anti-holomorphic automorphism is locally (i.e. in
local coordinates on every patch $V_{\nu}$) just ordinary complex
conjugation.
\end{proof}
$q$ satisfies all the demands in the interior of $R$. The double
of a curve $\gamma$ homotopic to a puncture on the boundary of
$R$, whose residue is $p$, is a closed curve $\tilde{\gamma}$
homotopic to the puncture on $\Gamma\in\hat{R}$ (this statement
holds individually for every connected component of $\Gamma$).
$\frac{1}{2\pi i}\oint_{\tilde{\gamma}}\sqrt{q}=p$ by virtue of
Theorem \ref{type1}. By symmetry $\frac{1}{\pi
i}\oint_{\gamma}\sqrt{q}=p$. The restriction of $q$ to $R$ is the
required differential. Uniqueness follows by considering a second
differential $q_{2}$ on $R$ which also meets the requirements of
the theorem. We can extend $q_{2}$ uniquely to the doubled surface
$\hat{R}$ by defining (in local coordinates) $q_2(z)\equiv {\bar
q}_2(\bar{z})$ where $\exists r\in R | h(r)=z$ and an image point
$\tilde{r}\in R^{*} | h(\tilde{r})=\bar{z}$. The differential
$q_{2}$ satisfies all the demands of Theorem \ref{type1} in the
interior of both $R$ and $R^{*}$. By considering the doubled
curves described above we can show that the demands also hold for
punctures on $\Gamma$. By theorem \ref{type1} then $q\equiv q_{2}$
and their respective restrictions to $R$ must also be equal.
\end{proof}

\subsection{Image charge method for open and closed worldsheets}

With this in hand we proceed to describe our method for
constructing Strebel differentials on Riemann surfaces with
boundary. We start with a field theory diagram (shown at the
beginning of each example). We interpret this diagram as a double
line graph as specified in \cite{'tHooft:1973jz}. We determine the
surface to which the diagram belongs, the borders and placement of
punctures. Both examples will consist of genus $0$ diagrams with
one boundary (disk diagrams). We construct the appropriate Strebel
differential using the familiar image charge method. First we
double the surface obtained from the field theory diagram. Every
operator insertion in the interior of the diagram gets a dual
image insertion in the doubled surface while insertions on the
boundary are left untouched. Now we construct the unique Strebel
differential for the boundary-less surface obtained. The details
of this construction may be found in
\cite{Gopakumar:2005fx,Aharony:2006th,Penkava} and in each
example. The resulting differential will have the property
$\phi(\bar{z})=\bar{\phi}(z)$ which is a reflection (no pun
intended) of the fact that we have placed image charges for each
interior insertion. We then restrict our differential to the
closed upper half plane which, for our diagrams, represents the
interior. By restrict we mean that all integrations and parameters
will take into account the fact that the metric is defined only on
the upper half plane. For example: the location for the zeros
$k_{i}$ of the function $\phi(z)$ will explicitly satisfy
$\Im(k)>0$. We identify the Schwinger parameters $\sigma_{i}$ with
the Strebel lengths $l_{i}$. Note that to apply Gopakumar's
prescription correctly we must identify the conductance of
boundary edges with the length of critical curves only up to the
boundary of the Riemann surface. This is made simpler by the image
charge method which guarantees that this is exactly half the
length of the full curve. A similar procedure may be used for
diagrams with more boundaries, though we will not analyze any
examples here.

We conclude by verifying that the moduli count on the two sides of
the correspondence is still the same. The Riemann-Roch theorem
states that: $m-k=-3\chi$ where $m$ is the number of real moduli of
the Riemann surface, $k$ is the real dimension of the conformal
Killing group (the number of conformal Killing vectors) and $\chi$
is the Euler characteristic. For a Riemann surface $R$ with boundary
$\chi=2-2g-b$ where $g$ is the genus of the compact surface from
which we cut $b$ holes to obtain $R$. If we add the positions of
$n_{c}$ closed string insertions living on the interior of $R$ and
$n_{o}$ open string insertions living on the boundary we get
$6g+3b-6+2n_{c}+n_{o}$ moduli (we assume that we have saturated the
space with enough insertions to account for all Killing vectors).
The decorated moduli space ${\cal M}_{g,b,n}\times
\mathbb{R}_{+}^{n}$ (b counts the boundaries) therefore has
$6g+3b-6+3n_{c}+2n_{o}$ moduli. It is easy to show that this is also
the number of moduli of a maximally connected ribbon graph with
genus $g$, $b$ boundaries and $n_{c}+n_{o}$ vertices of which
$n_{o}$ separate a boundary. To do this start with a maximally
connected genus $g$ graph and no boundaries with $n=n_{c}+n_{o}$
vertices. This has $6g-6+3n$ edges. Remove one face to create one
boundary. This has the effect of putting $b=1$, $n_{c}\rightarrow
n_{c}-3$ and $n_{o}\rightarrow n_{o}+3$ and does not change the line
count. This agrees with our formula for the Riemann surface.
Widening a boundary by deleting an adjacent face has the effect
$b\rightarrow b$, $n_{c}\rightarrow n_{c}-1$ and $n_{o}\rightarrow
n_{o}+1$ and we have deleted the edge separating the faces. This
also agrees with the Riemann surface calculation. Finally we may
split an internal line to create a boundary with only two vertices:
$b\rightarrow b+1$, $n_{c}\rightarrow n_{c}-2$ and $n_{o}\rightarrow
n_{o}+2$ which also fits (since there is now an additional edge).
Using these procedures we can recover any punctured Riemann surface
with boundary from the punctured compact Riemann surface of the same
genus.

\section{A three-point function example}
\label{three}

\begin{figure}
\centering
\includegraphics[height=4.67cm, keepaspectratio =
true]{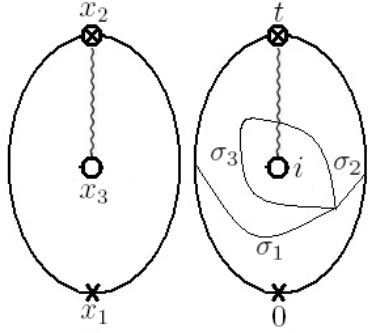}
\caption{\label{fig:threepointQFT}\footnotesize{The field theory
diagram and the dual graph. The solid line includes a propagator in
the fundamental representation which becomes a boundary on the
worldsheet. Big $X$'s represent $\bar{\Psi}\Psi$ insertions and
circles represent $\Phi$ insertions.}}
\end{figure}

We use Gopakumar's prescription to map the correlator shown in
figure \ref{fig:threepointQFT} to string theory. This correlator can
arise from the correlation function
$\langle\bar{\Psi}\Psi\left(x_{1}\right)
\bar{\Psi}\Phi\Psi\left(x_{2}\right)tr\left(\Phi\left(x_{3}\right)\right)\rangle$
in a $U(N)$ gauge theory, where $\Psi$ ($\bar{\Psi}$) are
(anti)-fundamental fields and $\Phi$ is in the adjoint
representation, or (after gluing) from more complicated diagrams
which we will analyze below. The double line graph describes a disk
with two boundary insertions and one interior insertion. Consulting
the dual graph (drawn on the right of figure
\ref{fig:threepointQFT}) we see that the appropriate differential
must have a single order $2$ zero in the interior of the surface.
Unlike the three point function described in \cite{Aharony:2006th}
this does not result in a trivial differential. This is because we
are working on a Riemann surface with boundary which restricts our
conformal Killing group to the one preserving the boundary. In this
case the group is $SL\left(2,\mathbb{R}\right)$ which is the
subgroup of the full conformal group of the sphere,
$SL\left(2,\mathbb{C}\right)$, which preserves our boundary: the
real line. We can (actually must) use this symmetry to fix the
positions of some of the insertions. $SL\left(2,\mathbb{R}\right)$
has dimension $3$ and we will use it to fix completely the position
of the interior insertion (two degrees of freedom) and of one of the
boundary insertions (one d.o.f. each). This leaves one unfixed
boundary operator whose position is integrated over. In particular
it can approach the other boundary insertion to generate an OPE.
notice that this choice of conformal frame is not unique, but the
number of unfixed degrees of freedom is. This will be discussed
further in section \ref{sec:Discussion}.

\begin{figure}
\centering
\includegraphics[height=8.2cm,keepaspectratio = true]{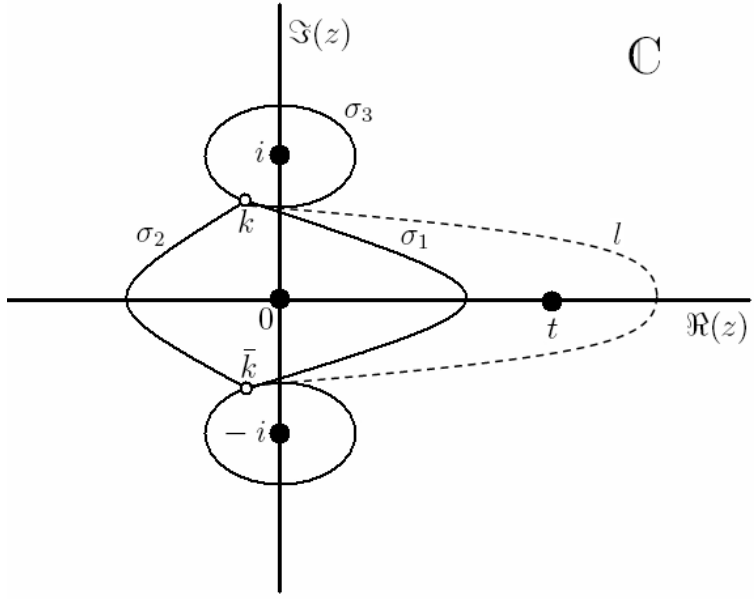}\caption{\label{fig:3pointStrebel}\footnotesize{The
critical graph of the Strebel differential. The solid balls mark
the positions of insertions. The small circles mark the positions
of the zeros of the differential.}}
\end{figure}

We fix the interior insertion at $i$ and the boundary insertion at
$0$. We choose the boundary insertion fixed at $0$ to be the one
not connected to the interior vertex. The position of the
remaining boundary insertion will be denoted by $t$. A general
quadratic differential of the correct form with a double zero is
given by
\begin{equation}
q=\phi(dz)^2=\frac{a (z-k)^2 \left(z-\bar{k}\right)^2}{z^2 (z-i)^2
(z+i)^2 (z-t)^2}(dz)^2.
\end{equation}
We demand $\Im(k)>0$ in all expressions. Notice that we have
placed an image charge at $-i$ and that $q$ has the symmetry
$\phi(\bar{z})=\bar{\phi}(z)$. We label the residues according to
their positions by $p_{z}$ and demand: $p_{i}=p_{-i}$. The fact
that the dual graph has an order $4$ vertex, which is the same
thing as saying that the differential has a double zero, imposes
an additional constraint on the residues: $p_{t}=p_{0}+2p_{i}$.
The differential in terms of these residues is:
\begin{equation}q=\left[\frac{i}{2\pi}\frac{\left(p_{0}t+2p_{i}t\right)z^2+\left(2p_{i}\right)z+p_{0}t}{z
(z-i) (z+i) (z-t)}\right]^2(dz)^2.\end{equation}
There are double zeros at:
\begin{equation}k,\bar{k}=\frac{-p_{i}\pm\sqrt{p_{i}^2-p_{0}^2t^2-2p_{0}p_{i}t^2}}{p_{0}t+2p_{i}t},\end{equation}
only one of which is inside our area of interest $\Im(z)> 0$. Note
that in order to get the correct graph we need $k$ and $\bar{k}$ to
be a complex conjugate pair. This means restricting the integration
over the $p_z$ and $t$ to where
$\sqrt{p_{i}^2-p_{0}^2t^2-2p_{0}p_{i}t^2}$ is imaginary or
equivalently $p_{i}^2-p_{0}^2t^2-2p_{0}p_{i}t^2< 0$. The invariant
line element is:
\begin{equation}\sqrt{\phi}dz=\frac{i}{2\pi}\frac{\left(p_{0}t+2p_{i}t\right)z^2+\left(2p_{i}\right)z+p_{0}t}{z
(z-i) (z+i) (z-t)}dz,\end{equation}
which we will integrate to get the undetermined length in the
critical graph:
\begin{equation}\int\sqrt{\phi}dz=\frac{i}{2\pi} \left(-p_{0} \log (z)+(p_{0}+2 p_{i} )
   \log (z-t)-p_{i}  \log \left(z^2+1\right)\right).\end{equation}
The integrated length along a specific line between the two zeros
(drawn in figure \ref{fig:3pointStrebel}) is:
\begin{equation}l=\int_{k}^{\bar{k}}\sqrt{\phi}dz=\frac{p_{0}}{\pi}\arg(k)-\frac{p_{0}+2 p_{i}}{\pi}\arg(k-t)+\frac{p_{i}}{\pi}\arg(k^2+1).\end{equation}
Note that $l$ is not one of the Strebel lengths, but that all
Strebel lengths can be derived from $l$ and the residues $p_{z_{i}}$
(see figure \ref{fig:3pointStrebel}). Define the quotient
$x\equiv\frac{p_{i}}{p_{0}}$, and a few expressions:
\begin{enumerate}
    \item $A\equiv x - i {\sqrt{t^2 + 2t^2x - x^2}}$
   \item
   $B\equiv -x - t^2\left( 1 + 2x \right)  - i {\sqrt{t^2 + 2t^2x -
      x^2}}$
    \item
    $C\equiv x + t^2\left( 1 + 2x \right)  + i {\sqrt{t^2 + 2t^2x - x^2}}$
\end{enumerate}
so that :
\begin{equation}l=\frac{ip_{0}}{2\pi}\left[\left(\log A-\log \bar{A}\right)+
(2x+1)\left(\log B-\log \bar{B}\right)+ x\left(\log C-\log
\bar{C}\right)      \right].\end{equation}
Notice that the expression is always real in the region of
integration which is where the two zeros are a conjugate pair and
our symmetry $\phi(\bar{z})=\bar{\phi}(z)$ holds, implying $t^2 +
2t^2x - x^2\ge 0$.

Let us write the field theory expression for the amplitude. We use
position space Schwinger parameters $\sigma_{i}$ (assuming a single
contraction of massless scalar fields along each line), such that
the correlation function is given by
\begin{equation}G(x_{1},x_{2},x_{3})=(const)\int_{0}^{\infty}\prod_{i=1}^{3}d\sigma_{i}
e^{-(\sigma_{1}+\sigma_{2})(x_{1}-x_{2})^2-\sigma_{3}(x_{2}-x_{3})^2}.\end{equation}

Finally, we are ready to insert the data from the critical graph of
the Strebel differential into this equation by making the
identification: $\text{(Strebel length)}\equiv\sigma_{i}$. Let us
write the appropriate dictionary in terms of our worldsheet moduli
(see figure \ref{fig:3pointStrebel} for the edges corresponding to
each $\sigma_i$) :

\begin{equation}\sigma_{1}=\frac{p_{0}+2p_{i}-l}{2},
    \qquad \sigma_{2}=\frac{l-2p_{i}}{2},
    \qquad \sigma_{3}=p_{i}.\end{equation}

We now change variables from $\sigma_{i}$ to
$\left(p_{0},x,t\right)$. The determinant of the transformation
matrix (the Jacobian) is:

\begin{equation}\textit{J}=\left|\frac{\partial\left(\sigma_{1},\sigma_{2},\sigma_{3}\right)}{\partial\left(p_{0},x,t\right)}\right|=
\left|\frac{\partial\left(\sigma_{1},\sigma_{2},\sigma_{3}\right)}{\partial\left(p_{0},p_{i},t\right)}\frac{\partial\left(p_{0},p_{i},t\right)}{\partial\left(p_{0},x,t\right)}\right|=\frac{p_{0}^2{\sqrt{t^2
+ 2t^2x - x^2}}}
  {4\pi t + 4\pi t^3}\geq 0.\end{equation}

We will integrate only over $t>0$ and multiply the result by $2$.
The correlator is:
\begin{align}
G(x_{1},x_{2},x_{3})=&2(const)\int d t d p_{0} d x
\frac{p_{0}^2{\sqrt{t^2 + 2t^2x - x^2}}}
  {4\pi t + 4\pi t^3}e^{
-p_{0}\left(\frac{1}{2}(x_{1}-x_{2})^2+x(x_{2}-x_{3})^2 \right),}
   \end{align}
where the bounds of integration are:
\begin{equation}p_{0}\in(0,\infty),
    \qquad t\in(0,\infty),
    \qquad x\in(0,t^2+\sqrt{t^2+t^4}).\end{equation}
Performing the integration over $p_{0}$ gives:
\begin{align}
G(x_{1},x_{2},x_{3})=&32(const)\int d t  d x \frac{{\sqrt{t^2 +
2t^2x - x^2}}}
  {4\pi t + 4\pi t^3}\frac{1}{((x_{1}-x_{2})^2+2x(x_{2}-x_{3})^2)^3.}
   \end{align}
We can now do the final integration over $x$. Define:
\begin{equation}K_{123}\equiv\frac{\left(x_{3}-x_{2}\right)^2}{\left(x_{2}-x_{1}\right)^2}\end{equation}
\begin{equation}r\equiv
2(K_{123}-1)\left(t^2+\sqrt{t^4+t^2}\right)-1,\qquad q\equiv
2K_{123}\left(t^2+\sqrt{t^4+t^2}\right)+1\end{equation}
then
\begin{equation}
\label{3pointfinal}G(x_{1},x_{2},x_{3})=\frac{(const)}{\left(x_{1}-x_{2}\right)^6}\int_{0}^{\infty}
d t\frac{(q-r-2) \left(\sqrt{q r} (q+r)-(q-r)^2 \tanh
^{-1}\left(\sqrt{\frac{r}{q}}\right)\right)}{\pi  (q r)^{3/2}
   \left(t^2+1\right)}\end{equation}
which is our final expression of the correlator as a function of the
modulus $t$. We do not have any particular insight into the meaning
of this expression.

Next, in order to compute the OPE of the two open string vertices we
expand around $t=0$. The first few terms of the integrand (ignoring
the $(const)$ from the correlator) are:
\begin{equation}G(x_{1},x_{2},x_{3})=\frac{2}{\left(x_{2}-x_{1}\right)^6}\,t +  \frac{\left( 8-16\,K_{123} \right)}{\pi \left(x_{2}-x_{1}\right)^6}t^2 + \frac{12\,\left( K_{123}-1 \right) \,K_{123}}{\left(x_{2}-x_{1}\right)^6}\,t^3 +
  {\text{O}(t)^4}.\end{equation}
Note that the power series involves integer powers of $t$,
suggesting operators of dimension $\Delta=3,4,...$ appearing in
the OPE. Note also that, with the exception of the leading term,
all odd orders in $t$ vanish at $K_{123}=1$ ($r=-1$) and all even
orders at $K_{123}=\frac{1}{2}$ ($r=-q$). Also the expression is
non-singular in $K_{123}$ so only positive powers of $K_{123}$
appear in the expansion. For large $K_{123}$ the $N'th$ order in
$t$ seems to scale as $K_{123}^{N-1}$ (for all N not just the ones
shown). The $K_{123}\rightarrow 0$ limit, which should be singular
if we consider the full correlator, is actually regular for each
term in the $t$ expansion, and the divergence comes from summing
the series.

Integrating the complete answer on the range $t\in(0,\infty)$ we of
course recover the expected correlator:
\begin{equation}G(x_{1},x_{2},x_{3})=\frac{(const)}{K_{123}\left(x_{1}-x_{2}\right)^6}=(const)\frac{1}{\left(x_{2}-x_{1}\right)^4\left(x_{3}-x_{2}\right)^2}\end{equation}

Next, we wish to consider more general correlators such as
\begin{equation}\label{general3point}G(x_{1},x_{2},x_{3})=\left\langle
\bar{\Psi}\Phi^{n_{1}}\Psi{\left(x_{1}\right)}
\bar{\Psi}\Phi^{n_{2}}\Psi{\left(x_{2}\right)}
tr\left(\Phi^{n_{3}}{\left(x_{3}\right)}\right)
\right\rangle\end{equation}
(with $n_2=n_1+n_3$) which also get
contributions from the same skeleton graph. In order to analyze
the OPE in this more general case we analyze the leading order
$t$-dependence of the various lines in the diagram. We change
variables to $y \equiv \frac{x}{t^2+\sqrt{t^2+t^4}}$. The range of
$y$ is from $0$ to $1$. The three lines are now:
\begin{enumerate}
    \item $\sigma_{1}=\frac{p_{0}}{2}+O(t)$
    \item $\sigma_{2}=\frac{p_{0}}{4\pi}\left(2\sqrt{1-y^2}-2y \arccos\left(y\right)\right)t +O(t^2)$
    \item $\sigma_{3}=p_{0}yt +O(t^2)$
\end{enumerate}
Adding lines to the diagram, by adding $\Phi$'s to the operators in
our field theory correlation function, changes the leading order $t$
dependence. An additional line homotopic to $\sigma_{2}$ or
$\sigma_{3}$ will increase the leading power by $1$. Additional
lines homotopic to $\sigma_{1}$, which is the line separating the
converging insertions, do not change the leading order behavior. For
example, the leading term after adding a $\sigma_{2}$ line is:
\begin{equation}G(x_{1},x_{2},x_{3})\approx\frac{64-12\pi}{3\pi^{2}\left(x_{2}-x_{1}\right)^{8}}t^2.\end{equation}

We can write down the result for the leading power appearing in the
OPE as a function of the number of adjoint contractions between the
various points. Let $L_{1}$ be the number of adjoint contractions
between $x_{1}$ and $x_{2}$ on the OPE side, and $L_{2}$ on the
other side, and let $L_{3}$ be the number of adjoint contractions
between $x_{2}$ and $x_{3}$. Then, the leading order in the
$t\rightarrow 0$ OPE is:
\begin{equation}G(x_{1},x_{2},x_{3})_{t\rightarrow 0}\sim t^{L_{2}+L_{3}}\end{equation}
We can also analyze the leading order behavior in the
$t\rightarrow 0$ limit as a function of the operators we insert in
the field theory correlator (\ref{general3point}) described above.
The two sets are related by:
\begin{equation}n_{1}=L_{1}+L_{2},
   \qquad n_{2}=L_{1}+L_{2}+L_{3},
   \qquad n_{3}=L_{3}.\end{equation}
The lowest order contribution to the OPE with given $\{n_{i}\}$
will come from loading all possible lines on the $\sigma_{1}$
side, taking $L_{2}=0$, giving
\begin{equation}G(x_{1},x_{2},x_{3})_{t\rightarrow 0}\sim t^{n_{3}},\end{equation}
meaning that the dimension of the leading operator contributing to
the OPE of the operators dual to $\bar{\Psi}\Phi^{n_{1}}\Psi$ and
$\bar{\Psi}\Phi^{n_{2}}\Psi$ in this diagram is
$n_{3}+2=n_{2}-n_{1}+2$.

\section{A four-point function example}
\label{four}

\begin{figure}
\centering
\includegraphics[height=4.57cm, keepaspectratio =
true]{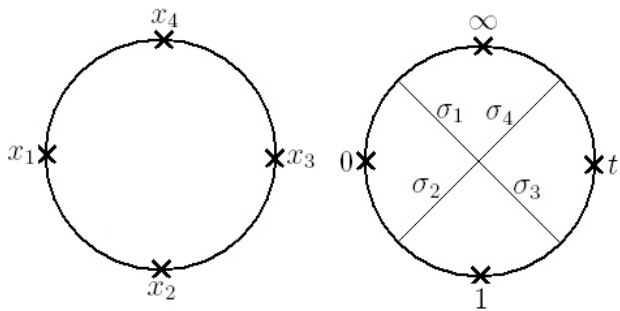}\caption{\label{fig:fourpointQFT}\footnotesize{The
field theory diagram and the dual graph of the 4 point correlator.
The four solid lines form the boundary of a disk.}}
\end{figure}

In this section we evaluate the four point correlator shown in
figure \ref{fig:fourpointQFT}, which corresponds in the field theory
to a correlator such as
$\langle\bar{\Psi}\Psi\left(x_{1}\right)\bar{\Psi}\Psi\left(x_{2}\right)\bar{\Psi}\Psi\left(x_{3}\right)\bar{\Psi}\Psi\left(x_{4}\right)\rangle$.
The dual line graph describes a disk with four boundary insertions.
Our conformal Killing group is again $SL\left(2,\mathbb{R}\right)$.
We use this symmetry to fix the positions of three of the boundary
insertions at $0$, $1$ and $\infty$. Notice that the fixing does not
change the cyclic order of the insertions, which has to be summed
over to compute the full correlator.

Again, the dual graph implies that the Strebel differential must
have a double zero. A general quadratic differential of the correct
form with a double zero is :
\begin{equation}q=\phi(dz)^2=\frac{a (z-k)^2 (z-\bar{k})^2}{z^2 (z-1)^2 (z-t)^2}(dz)^2.\end{equation}
We have one relation between the residues:
\begin{equation}p_{0}+p_{t}=p_{\infty}+p_{1},\end{equation}
where we have assumed a cyclic order such that $t>1$. The
differential in terms of the residues is:
\begin{equation}q=\left[\frac{1}{2 \pi i}\frac{p_{\infty} z^2+(-p_{\infty}+
p_{1} (t-1)-p_{0} t) z+p_{0} t}{z (z-1)
(z-t)}\right]^2(dz)^2,\end{equation}
which has double zeros at:
\begin{equation}k,\bar{k}=\frac{1}{2p_{\infty}}\left( p_{1}+p_{\infty}+
t(p_{0}-p_{1})\pm \sqrt{
(p_{1}+p_{\infty}+t(p_{0}-p_{1}))^2-4p_{0}p_{\infty}t      }
\right)\end{equation}
To find the range of integration we demand that $\Im(k) > 0$ and the
two zeros form a conjugate pair. This gives the constraint:
\begin{equation}(p_{1}+p_{\infty}+t(p_{0}-p_{1}))^2-4p_{0}p_{\infty}t\leq 0,\end{equation}
or equivalently:
\begin{equation}t-x+t x-2\sqrt{t^2 x - t x}\leq y \leq t-x+t x+2\sqrt{t^2 x - t
x},\end{equation}
where we have defined the ratios
$x\equiv\frac{p_{1}}{p_{0}},y\equiv\frac{p_{\infty}}{p_{0}}$.

\begin{figure}
\centering
\includegraphics[height=7.9cm, keepaspectratio =
true]{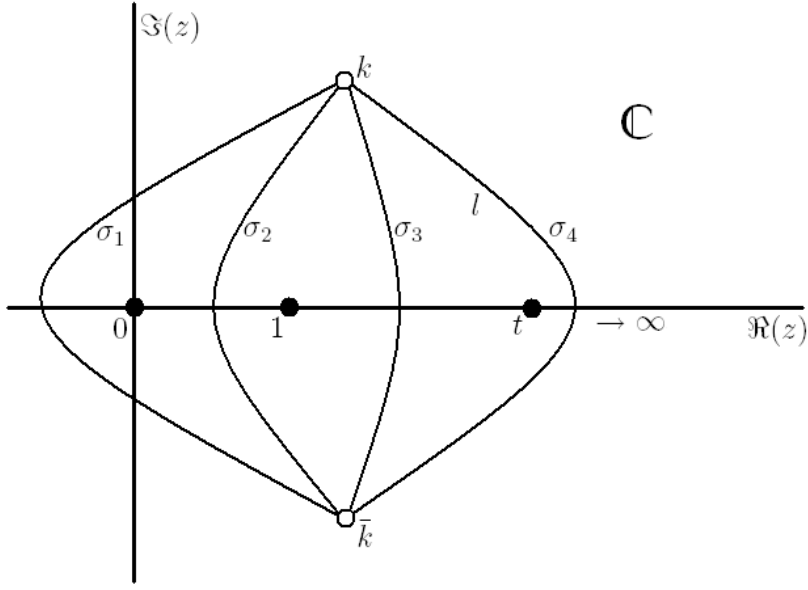}\caption{\label{fig:4pointStrebel}\footnotesize{The
critical graph of the 4-point Strebel differential.}}
\end{figure}

The invariant line element is:
\begin{equation}\sqrt{\phi} dz=\frac{1}{2 \pi i}\frac{p_{\infty} z^2+
(-p_{\infty}+p_{1} (t-1)-p_{0} t) z+p_{0} t}{z (z-1)
(z-t)}dz,\end{equation} and the integrated metric is :
\begin{equation}\int\sqrt{\phi} dz=\frac{1}{2\pi i}(p_{0}\log
   (z)-p_{1}\log
(z-1)+\left(p_{\infty}+p_{1}-p_{0}\right) \log
(z-t)).\end{equation}
The integrated length along one of the edges in the graph (see
figure \ref{fig:4pointStrebel}) is:
\begin{equation}l=\int_{k}^{\bar{k}}\sqrt{\phi}dz=\frac{p_{0}}{\pi}\arg(k)-
\frac{p_{1}}{\pi}\arg(k-1)+\frac{\left(p_{\infty}+p_{1}-p_{0}\right)}{\pi}\arg(k-t).\end{equation}
We define a few more functions:
\begin{enumerate}
    \item
    $\tilde{A}\equiv t + x - tx + y + {\sqrt{-4ty + {\left( t + x - tx + y \right)
}^2}},$
   \item
   $\tilde{B}\equiv t + x - tx - y - {\sqrt{-4ty + {\left( t + x - tx + y \right) }^2}},$
    \item
   $\tilde{C}\equiv t + x - tx + y - 2ty + {\sqrt{-4ty + {\left( t + x - tx + y \right) }^2}},$
\end{enumerate}
such that :
\begin{equation}l=\frac{ip_{0}}{2\pi}\left[\left(\log \tilde{A}-\log \bar{\tilde{A}}\right)-x
\left(\log \tilde{B}-\log \bar{\tilde{B}}\right)+
(x+y-1)\left(\log \tilde{C}-\log \bar{\tilde{C}}\right)
\right]\end{equation}
Note that this expression is real in the region of integration.

Next, we construct the dictionary between the $\sigma_{i}$'s and the
worldsheet moduli for the regime where
$p_{0}+p_{t}=p_{\infty}+p_{1}$:
\begin{equation}\sigma_{1}=\frac{1}{2}(p_{\infty}-l),
    \qquad \sigma_{2}=\frac{1}{2}(p_{0}-p_{\infty}+l),
    \qquad \sigma_{3}=\frac{1}{2}(p_{1}+p_{\infty}-p_{0}-l),
   \qquad \sigma_{4}=\frac{l}{2}.\end{equation}
The range of integration is
\begin{equation}t\in(1,\infty),
    \qquad p_{0}\in(0,\infty),
     \qquad x\in(0,\infty),$$
     $$ y\in\left(t - x + t x - 2{\sqrt{t^2x -tx   }},t - x + t x + 2{\sqrt{  t^2x-tx
    }}\right),\end{equation}
where we have taken into account the support for the
$\delta$-function integration
$\delta\left(p_{t}-\left(p_{\infty}+p_{1}-p_{0}\right)\right)$. The
Jacobian for the change of variables from
$\left(\sigma_{1},\sigma_{2},\sigma_{3},\sigma_{4}\right)$ to
$\left(p_{0},x,y,t\right)$ is:
\begin{align}
\textit{J}=\frac{-ip_{0}^3}{32}\frac{
    {\sqrt{
        {\left( t + x - t\,x + y \right) }^2 -4ty }}}{\pi \,
    \left( t-1\right) \,t}.
\end{align}
Note that $\textit{J}$ is real and positive in the range of
integration. Once again we write the field theory expression for the
amplitude. Define also $J_{1}$ as the jacobian for the change of
variables from
$\left(\sigma_{1},\sigma_{2},\sigma_{3},\sigma_{4}\right)$ to
$\left(p_{0},p_{1},p_{\infty},t\right)$. The correlator in position
space is given by:
\begin{align}
\label{fourpointcorrelator}
&G(x_{1},x_{2},x_{3},x_{4})=\int_{0}^{\infty}\prod_{i=1}^{4}d\sigma_{i}e^{-\sigma_{1}\left(x_{4}-x_{1}\right)^{2}-\sigma_{2}\left(x_{1}-x_{2}\right)^{2}-\sigma_{3}\left(x_{2}-x_{3}\right)^{2}-\sigma_{4}\left(x_{3}-x_{4}\right)^{2}}
\\&=\int dt dp_{0} dp_{1} dp_{\infty}\textit{J}_{1}e^{-\frac{1}{2}\left(p_{\infty}-l\right)\left(x_{4}-x_{1}\right)^{2}-\frac{1}{2}(p_{0}-p_{\infty}+l)\left(x_{1}-x_{2}\right)^{2}-\frac{1}{2}(p_{1}+p_{\infty}-p_{0}-l)\left(x_{2}-x_{3}\right)^{2}-\frac{l}{2}\left(x_{3}-x_{4}\right)^{2}}
\\&=\int dt dp_{0} dx
dy\textit{J}e^{-\frac{1}{2}\left(yp_{0}-l\right)\left(x_{4}-x_{1}\right)^{2}-\frac{1}{2}(p_{0}-yp_{0}+l)\left(x_{1}-x_{2}\right)^{2}-\frac{1}{2}(xp_{0}+yp_{0}-p_{0}-l)\left(x_{2}-x_{3}\right)^{2}-\frac{l}{2}\left(x_{3}-x_{4}\right)^{2}}.
\end{align}
We perform the $dp_{0}$ integration (which kills the exponent):
\begin{align}
&G(x_{1},x_{2},x_{3},x_{4})=\int dt dx dy \frac{-3i}{16\pi}\frac{
    {\sqrt{
        {\left( t + x - t\,x + y \right) }^2 -4ty }}}{ t^2-t}
\\&
\left(\frac{y-\tilde{l}}{2}\left(x_{4}-x_{1}\right)^{2}+\frac{1-y+\tilde{l}}{2}\left(x_{1}-x_{2}\right)^{2}+\frac{x+y-1-\tilde{l}}{2}\left(x_{2}-x_{3}\right)^{2}+\frac{\tilde{l}}{2}\left(x_{3}-x_{4}\right)^{2}\right)^{-4},
\end{align}
where $\tilde{l}\equiv \frac{l}{p_{0}}$ does not depend on $p_{0}$.

We were not able to do the explicit integral over $x$ and $y$ to
obtain the full correlation function, but we can analyze it in the
two OPE limits $t\rightarrow 1$ and $t\rightarrow \infty$. We switch
to the variables:
 \begin{equation}\begin{cases}
 t\rightarrow 1& u\equiv\frac{ t \left( x-1 \right)  -
      x - y  }{{\sqrt
       {ty}}},\\
       t\rightarrow \infty& u\equiv\frac{y-t+x-tx}{\sqrt{t^2x-tx}}.
        \end{cases}\end{equation}
The ranges of integration are now
\begin{equation}t\in(1,\infty)
    \qquad x,y\in(0,\infty)
    \qquad u\in\left(-2,2\right)\end{equation}
The leading order contributions to the $dt$ integral are now:
\begin{equation}G(x_{1},x_{2},x_{3},x_{4})\approx\begin{cases}
t\rightarrow 1& \int dy du \frac{\left( -3i  \right)
    {\sqrt{-4 + u^2}}y}{\pi {\left( 1 +
        u{\sqrt{y}} + y \right)
        }^4{\left( x_{3} - x_{2}
        \right) }^8}    \left(t-1\right)^2=\frac{3}{8{\left( x_{3} - x_{2}
        \right) }^8}\left(t-1\right)^2,\\
t\rightarrow \infty& \int dx du\frac{\left( -3i  \right)
    {\sqrt{-4 + u^2}}x}{\pi {\left( 1 +
        u{\sqrt{x}} + x \right)
        }^4{\left( x_{3} - x_{4}
        \right) }^8}    \frac{1}{t^4}=\frac{3}{8{\left( x_{3} - x_{4}
        \right) }^8}\frac{1}{t^4}.
\end{cases}\end{equation}
Note the similarity between these expressions. The two OPE limits
are of course identical, up to exchanging of the $x_{i}$,  from
the field theory point of view. On the worldsheet the two limits
are connected by the transformation $t\rightarrow\frac{t}{t-1}$,
which exchanges $1$ and $\infty$ and keeps $0$ fixed. The
difference in the power of the leading order stems from the
Jacobian of this transformation, $\frac{1}{(t-1)^2}$. The result
suggests that the OPE expansion contains operators of worldsheet
dimension $\Delta=4,5,..$. We have checked numerically that the
final integrated answer scales as: (as is obvious from
(\ref{fourpointcorrelator}))
\begin{equation}(const)\frac{1}{\left(x_{2}-x_{1}\right)^2\left(x_{3}-x_{2}\right)^2\left(x_{4}-x_{3}\right)^2\left(x_{1}-x_{4}\right)^2}.\end{equation}

We can add additional adjoint lines to this correlator, computing
more general correlators of the form
\begin{equation}G(x_{1},x_{2},x_{3},x_{4})=\left\langle 0 \left|
\bar{\Psi}\Phi^{n_{1}}\Psi{\left(x_{1}\right)}
\bar{\Psi}\Phi^{n_{2}}\Psi{\left(x_{2}\right)}
\bar{\Psi}\Phi^{n_{3}}\Psi{\left(x_{3}\right)}
\bar{\Psi}\Phi^{n_{4}}\Psi{\left(x_{4}\right)}
\right|0\right\rangle.\end{equation}
Note that there are additional diagrams that contribute to this
correlator, but we will only be interested in those Feynman diagrams
which upon gluing give the diagram of figure \ref{fig:fourpointQFT}.
Even in this diagram there are different ways to distribute the
contractions for given operators. In order to analyze the general
OPE we need to compute the scalings of the various edges in the OPE
limits, which are given in Table \ref{scalings}.

\begin{table}
\centering
\begin{tabular}{|c||c|c|c|c|} \hline
edge & connects & worldsheet value & $t\rightarrow1$ & $t\rightarrow\infty$ \\
\hline\hline
$\sigma_{1}$ & $x_{4}\longleftrightarrow x_{1}$ & $\frac{1}{2}(p_{\infty}-l)$ & $O\left(t-1\right)$ & $O\left(\frac{1}{t}\right)$ \\
\hline
$\sigma_{2}$ & $x_{1}\longleftrightarrow x_{2}$ & $\frac{1}{2}(p_{0}-p_{\infty}+l)$ & $O\left(t-1\right)$ & $O\left(\frac{1}{t}\right)$ \\
\hline
$\sigma_{3}$ & $x_{2}\longleftrightarrow x_{3}$ & $\frac{1}{2}(p_{1}+p_{\infty}-p_{0}-l)$ & $O\left(1\right)$ & $O\left(\frac{1}{t}\right)$ \\
\hline
$\sigma_{4}$ & $x_{3}\longleftrightarrow x_{4}$ & $\frac{1}{2}l$ & $O\left(t-1\right)$ & $O\left(1\right)$ \\
\hline
\end{tabular}
\caption{\label{scalings}Edge scaling in the two OPE limits.}
\end{table}

Let $L_{i}$ be the number of adjoint contractions between $x_{i}$
and $x_{i+1}$ ($L_{4}$ is between $x_{4}$ and $x_{1}$). Then, the
leading orders in the OPE's are:
\begin{equation}G(x_{1},x_{2},x_{3},x_{4})\sim\begin{cases}
t\rightarrow 1& t^{L_{1}+L_{3}+L_{4}+2},\\
t\rightarrow \infty&
\left(\frac{1}{t}\right)^{L_{1}+L_{2}+L_{4}+4}.
\end{cases}\end{equation}
In the correlator described above we have :
\begin{equation}n_{1}=L_{4}+L_{1},
    \qquad n_{2}=L_{1}+L_{2},
    \qquad  n_{3}=L_{2}+L_{3},
    \qquad  n_{4}=L_{3}+L_{4}.\end{equation}
These must satisfy $n_{1}+n_{3}=n_{2}+n_{4}$ in order to connect
all lines. Given this constraint it is always possible to connect
at most $\min\left(n_{2},n_{3}\right)$ lines between $x_{2}$ and
$x_{3}$, or $\min\left(n_{3},n_{4}\right)$ lines between $x_{3}$
and $x_{4}$. These contractions will contribute to the leading
order of the OPE's:
\begin{equation}G(x_{1},x_{2},x_{3},x_{4})\sim\begin{cases}
t\rightarrow 1& \left(t-1\right)^{\frac{1}{2}\left(n_{1}+n_{4}+|n_{2}-n_{3}|\right)+2}=\left(t-1\right)^{\max\left(n_{1},n_{4}\right)+2},\\
t\rightarrow \infty&
\left(\frac{1}{t}\right)^{\max\left(n_{1},n_{2}\right)+4}.
\end{cases}\end{equation}
Note that the limit $t\rightarrow 1$ corresponds to the OPE between
the vertex operators corresponding to
$\bar{\Psi}\Phi^{n_{2}}\Psi{\left(x_{2}\right)}$ and
$\bar{\Psi}\Phi^{n_{3}}\Psi{\left(x_{3}\right)}$, while the limit
$t\rightarrow\infty$ corresponds to the OPE between the vertex
operators corresponding to
$\bar{\Psi}\Phi^{n_{3}}\Psi{\left(x_{3}\right)}$ and
$\bar{\Psi}\Phi^{n_{4}}\Psi{\left(x_{4}\right)}$.

\section{Discussion and conclusions}
\label{sec:Discussion}

As was discussed in \cite{Aharony:2006th}, the worldsheet
expressions obtained from this formalism do not realize the
space-time conformal symmetry as a local symmetry. In our open
string diagrams we can see this already at the level of the
3-point function (\ref{3pointfinal}). Of the possible space-time
transformations, only the Poincar\'{e} and scaling symmetries are
locally realized global symmetries of the integrands in the
examples. The full conformal symmetry is of course restored after
we integrate over all the moduli.

In both examples the leading order contribution to an OPE depended
on the multiplicities of all lines not connecting the converging
operators. At first sight this is surprising since the OPE should
depend on the operators which are converging. However, this can be
seen as a manifestation of worldsheet conformal invariance. Consider
the convergence of $2$ out of $n$ operators on the boundary (which
is present for planar diagrams) or in the interior of a worldsheet.
If we assume a large enough conformal group then as the operators
come together we may always choose a conformal frame where the
converging operators are held fixed, say at $0$ and $1$. It is easy
to show that in this frame the rest of the operators converge on
each other. The two limits, the one of two operators converging and
the one of $n-2$, are thus equivalent. It is not surprising that the
OPE computed through this $n$-point diagram will depend on all the
multiplicities in the diagram.

The operator product expansions in both examples, and in the
general analysis of appendix \ref{circle}, consist of positive
integer powers of the separation. This suggests that the operators
appearing in the expansion also have positive integer worldsheet
dimensions (starting from dimension $3$ in the three point example
and dimension $n$ in the general circle diagram of order $n$).

\acknowledgments The work presented here was carried out under the
supervision of Ofer Aharony. We would like to thank Ofer Aharony and
Zohar Komargodski for careful reading of the manuscript and for
countless discussions. We gratefully acknowledge the help of Assaf
Patir, Dori Reichmann, Jacob Kagan and Yonatan Savir. This work was
supported in part by the Israel-U.S. Binational Science Foundation,
by the Israel Science Foundation (grant number 1399/04), by the
Braun-Roger-Siegl foundation, by the European network
HPRN-CT-2000-00122, by a grant from the G.I.F., the German-Israeli
Foundation for Scientific Research and Development, and by Minerva.

\appendix
\section{The general circle diagram of even order}
\label{circle}
 We can generalize the results of the four-point example of section
 \ref{four} to
determine the OPE's for the general circle diagram with an even
number $n$ of insertions. In these types of diagrams there is only
one unknown length to compute. The first fact we will use is that
the differential for a circle diagram with an even number of
insertions is a perfect square. We will use the same conventions
as in the four point example and set the position of one insertion
to $z_{1}=0$, another to $z_{2}=1$ and another to $z_{n}=\infty$,
and choose the rest of the insertions,
$\{z_{3},z_{4},..,z_{n-1}\}$ to lie between $1$ and $\infty$. The
invariant line element is now a rational function with a numerator
which has two conjugate zeros, $k$ and $\bar{k}$, each with
multiplicity $n/2-1$ and a denominator which is the product of
$n-1$ distinct monomials:
\begin{equation}\sqrt{\phi\left(z\right)}dz=\frac{p_{\infty}\left(z-k\right)^{\frac{n}{2}-1}
\left(z-\bar{k}\right)^{\frac{n}{2}-1}}{2\pi
i\prod_{j=1}^{n-1}\left(z-z_{j}\right)}dz.\end{equation}
This type of function can always be expressed as the sum of
rational functions where the denominators are the monomials and
the numerators are numbers. The numerators must be the residues
and the entire differential is so determined:
\begin{equation}\sqrt{\phi\left(z\right)}dz=\sum_{j=1}^{n-1}\frac{\left(-1\right)^{j+1}
p_{z_{j}}}{2\pi i} \frac{1}{z-z_{j}}dz.\end{equation}
Note the alternating signs needed to ensure that the single
residue relation,
\begin{equation}p_{\infty}=\sum_{i=1}^{n-1}(-1)^{i+1}
p_{z_{i}},\end{equation}
holds. The integrated length is just a sum of logarithms which is
equivalent to a sum of arguments of $k-z_{i}$ as we showed in the
example. The coefficients are just the residues (with alternating
signs):
\begin{equation}l=\int_{k}^{\bar{k}}\sqrt{\phi\left(z\right)}dz=\sum_{j=1}^{n-1}
\frac{\left(-1\right)^{j+1}p_{z_{j}}}{\pi}
\arg\left(k-z_{j}\right)\end{equation}
We denote the position of the insertion closest to $\infty$ by
$t$. We will analyze the scaling of the different components in
the calculation as a function of $t$. Notice the residue relation
$p_{t}=\sum_{j=1}^{n-2} (-1)^{j}p_{z_{j}}+p_{\infty}$ implies that
the factor $p_{\infty}$ appears only in the numerator of the final
fraction:
\begin{equation}\frac{\sum_{j=1}^{n-2} (-1)^{j}p_{z_{j}}+p_{\infty}}{2\pi i}\frac{1}{z-t}.\end{equation}
We take the $n$ parameters of the field theory to be the $n-1$
independent $p_{z_{i}}$ and the modulus $t$. All other position
moduli must then be determined as a function of these in order to
retain a differential with a single conjugate pair of zeros (one
that fits our diagram). Notice that the $t$ scaling of these
functions has no bearing on the scaling of the undetermined length
$l$. In fact there appears to be no parameter in $l$ that can scale
with $t$. This is not completely true since one must still restrict
the range of integration of one residue, which we take to be
$p_{\infty}$, so that the two zeros form a conjugate pair. Assume
that the functions that fix the remaining $z_{i}$, in terms of $t$
and the $p_{z_{i}}$, have been determined. Assume also that we have
performed a change of variables so as to fix the $t$-dependent
boundaries of integration for $p_{\infty}$, and replaced
$p_{\infty}$ with a variable $u$, which does not scale with $t$, and
some explicit dependence on $t$ and the other $p_{z_{i}}$. We
examine the $t\rightarrow\infty$ limit keeping $p_{z_{i}\neq\infty}$
and $u$ fixed. We assume that we can take this limit inside the
various integrations. The numerator for the invariant line element
is now a polynomial of the form
\begin{equation}P(z)=\left(A z^2+B z+C\right)^{\frac{n}{2}-1}.\end{equation}
Some of the coefficients are :
\begin{itemize}
\item $\text{coef.}(z^{n-2})=A^{\frac{n}{2}-1},$ \item
$\text{coef.}(1)= C^{\frac{n}{2}-1},$\item
$\text{coef.}(z)=\left(\frac{n}{2}-1\right)B C^{\frac{n}{2}-2}.$
\end{itemize}
We can write the same coefficients in terms of the $p_{z_{i}}$ and
the $z_{i}$ :
\begin{itemize}
\item $\text{coef.}(z^{n-2})=p_{\infty},$ \item
$\text{coef.}(1)=p_{0}\prod_{z_{i}\neq 0} z_{i},$ \item
$\text{coef.}(z)=\sum_{i=1}^{n-1}(-1)^{i+1}p_{z_{i}}\sum_{j\neq i}
\prod_{k\neq i,j} (-z_{k}).$
\end{itemize}
The condition for getting a pair of conjugate zeros is $B^2<4AC$,
which we viewed as an equation for the range of $p_{\infty}$. The
boundaries of the region of integration obey $B^2=4AC$, and we
require this to have (non-degenerate) solutions in the large $t$
limit. Denote by $\alpha_{i}$ the scaling of the (fixed) position
$z_{i}$ (so that, for $t\gg 1$, $z_{i}\sim t^{\alpha_{i}}$). We
demand $\alpha_{i}\geq 0\ \forall i$, otherwise two poles would
collide for some finite $t$. Denote by $p$ the scaling of
$p_{\infty}$, and by $(a,b,c)$ the scalings of $(A,B,C)$. We can now
write equations for the relationships between these scalings:
\begin{itemize}
\item $a=\frac{p}{\frac{n}{2}-1}$ \item
$c=\frac{1+\sum_{i}\alpha_{i}}{\frac{n}{2}-1}$ \item $B^2=4AC
\Rightarrow b=\frac{1}{2}(a+c)=\frac{1+p+\sum_{i}\alpha_{i}}{n-2}$
\item
$b+(\frac{n}{2}-2)c=max\{\sum_{i}\alpha_{i}+1,\sum_{i}\alpha_{i}+p
\}$
\end{itemize}
with the final equation coming from considering the largest possible
scaling for the coefficient of $z$. We can show that these imply
$p=1$, since :
\begin{equation}\begin{cases}
p>1& \Rightarrow \sum_{i}\alpha_{i}+p=\frac{1+p+\sum_{i}\alpha_{i}}{n-2}+(\frac{n}{2}-2)\left(\frac{1+\sum_{i}\alpha_{i}}{\frac{n}{2}-1}\right)\Rightarrow\\
&p=1-\frac{\sum_{i}\alpha_{i}}{n-3}\leq 1\ (contradiction), \\ p<1& \Rightarrow \sum_{i}\alpha_{i}+1=\frac{1+p+\sum_{i}\alpha_{i}}{n-2}+(\frac{n}{2}-2)\left(\frac{1+\sum_{i}\alpha_{i}}{\frac{n}{2}-1}\right)\Rightarrow\\
&p=1+\sum_{i}\alpha_{i}\geq 1\ (contradiction).
\end{cases}\end{equation}
This in turn implies $\alpha_{i}= 0\ \forall i$ and
$a=b=c=\frac{2}{n-2}$. The second equality tells us that the
positions of the zeros, $k$ and $\bar{k}$, do not scale with $t$.

Given this we are in exactly the same situation as we were for the
$4$ point example. For $t\rightarrow\infty$ we have $l\sim
p_{\infty}$. Note that for $t\gg 1$ $\frac{1}{\pi}\arg(k-t)\approx
1$. All other lines in the diagram are expressions of the type
$(p_{\infty}+f(\{p_{z}\}_{z\neq \infty}))-l$ and will therefore not
scale with $t$. The last thing to consider is the integration over
$p_{0}$. This kills the exponent and brings the expression
$-\sum_{s=1}^{n}\tilde{\sigma}_{s}\left(x_{s(1)}-x_{s(2)}\right)^2$,
$\tilde{\sigma}_{s}\equiv\sigma_{s}/p_{0}$ into the denominator. The
power to which this expression is raised is set by the dimension of
the correlator. The initial power is $n$ and adding one line adds
one power. All the $\sigma_{i}$, with the exception of the one
associated with $l$, contain $p_{\infty}-l$ and do not scale with
$t$. The denominator therefore scales as $t^{n}$. We can follow the
changes of variables to show that the product of the various
Jacobians does not scale with $t$. The leading order in the OPE with
$t\rightarrow\infty$ is therefore $\left(\frac{1}{t}\right)^{n}$
which is consistent with our analysis of the four point example.
Thus, the lowest dimension operator contributing to the OPE has
dimension $n$. Adding lines connecting the converging operators, the
line associated with $l$, does not change this as $l$ scales
linearly and the power to which the linearly scaling expression in
the denominator is raised increases by one. Adding any other line
decreases the leading power by one, as these lines do not scale with
$t$ but have the same effect on the denominator.

We can probably do a similar analysis for even circle correlators
with interior insertions connected to one of the boundary insertions
(like in the three-point example, but with more boundary and
interior insertions). There would be no restriction on the number of
interior insertions, as each contributes two to the orders of the
two zeros $k$ and $\bar{k}$. Other types of diagrams whose
differentials are perfect squares may also be possible to analyze in
general.

\bibliographystyle{amsplain}

\end{document}